\documentstyle[12pt,epsfig]{article}
\def \Et {{\rm E}_{\rm T}}
\def \Pt {{\rm P}_{\rm T}}

\def\Z0{${\em Z^0\/}$}

\def\r#1 {$^{#1}$}

\newcommand{\qqbar}{q\bar{q}}
\newcommand{\ttbar}{t\bar{t}}
\newcommand{\bbbar}{b\bar{b}}

\newcommand{\prl}[1]{Phys. Rev. Lett {\bf #1}}
\newcommand{\prev}[1]{Phys. Rev. {\bf #1}}

\newcommand{\prep}[1]{Phys. Rep. {\bf #1}}
\newcommand{\rmp}[1]{Rev. Mod. Phys. {\bf #1}}

\newcommand{\nim}[1]{Nucl. Instrumen. Meth. {\bf #1}}
\def\gepsfcentered#1{
  \def\testit{#1}
  \def\lbracket{[}
  \ifx\testit\lbracket
    \let\dofilecmd=\gepsfwithopt
  \else
    \let\dofilecmd=\gepsfnoopt
  \fi
  \dofilecmd}

\def\gepsfnoopt#1{
  \begin{center}
  \leavevmode
  \epsffile{#1}
  \end{center}}

\def\gepsfwithopt#1 #2 #3 #4]#5{
  \begin{center}
  \leavevmode
  \gepsfmaxx=0.94\textwidth
  \epsffile[#1 #2 #3 #4]{#5}
  \end{center}}

\newdimen\gepsfmaxx
\def\epsfsize#1#2{
  \ifnum \epsfxsize=0
    \ifnum \epsfysize=0
      \ifnum #1 > \gepsfmaxx
        \gepsfmaxx
      \else
        #1
      \fi
    \else
      \epsfxsize
    \fi
  \else
    \epsfxsize
  \fi
}

\def\be{\begin{equation}}
\def\ee{\end{equation}}
\def\bea{\begin{eqnarray}}
\def\eea{\end{eqnarray}}

\def\ap#1#2#3   {{\em Ann. Phys. (NY)} {\bf#1} (#2) #3.}
\def\apj#1#2#3  {{\em Astrophys. J.} {\bf#1} (#2) #3.}
\def\apjl#1#2#3 {{\em Astrophys. J. Lett.} {\bf#1} (#2) #3.}
\def\app#1#2#3  {{\em Acta. Phys. Pol.} {\bf#1} (#2) #3.}
\def\ar#1#2#3   {{\em Ann. Rev. Nucl. Part. Sci.} {\bf#1} (#2) #3.}
\def\cpc#1#2#3  {{\em Computer Phys. Comm.} {\bf#1} (#2) #3.}
\def\err#1#2#3  {{\it Erratum} {\bf#1} (#2) #3.}
\def\ib#1#2#3   {{\it ibid.} {\bf#1} (#2) #3.}
\def\jmp#1#2#3  {{\em J. Math. Phys.} {\bf#1} (#2) #3.}
\def\ijmp#1#2#3 {{\em Int. J. Mod. Phys.} {\bf#1} (#2) #3.}
\def\jetp#1#2#3 {{\em JETP Lett.} {\bf#1} (#2) #3.}
\def\jpg#1#2#3  {{\em J. Phys. G.} {\bf#1} (#2) #3.}
\def\mpl#1#2#3  {{\em Mod. Phys. Lett.} {\bf#1} (#2) #3.}
\def\nat#1#2#3  {{\em Nature (London)} {\bf#1} (#2) #3.}
\def\nc#1#2#3   {{\em Nuovo Cim.} {\bf#1} (#2) #3.}
\def\nim#1#2#3  {{\em Nucl. Instr. Meth.} {\bf#1} (#2) #3.}
\def\np#1#2#3   {{\em Nucl. Phys.} {\bf#1} (#2) #3.}
\def\pcps#1#2#3 {{\em Proc. Cam. Phil. Soc.} {\bf#1} (#2) #3.}
\def\pl#1#2#3   {{\em Phys. Lett.} {\bf#1} (#2) #3.}
\def\prep#1#2#3 {{\em Phys. Rep.} {\bf#1} (#2) #3.}
\def\prev#1#2#3 {{\em Phys. Rev.} {\bf#1} (#2) #3.}
\def\prl#1#2#3  {{\em Phys. Rev. Lett.} {\bf#1} (#2) #3.}
\def\prs#1#2#3  {{\em Proc. Roy. Soc.} {\bf#1} (#2) #3.}
\def\ptp#1#2#3  {{\em Prog. Th. Phys.} {\bf#1} (#2) #3.}
\def\ps#1#2#3   {{\em Physica Scripta} {\bf#1} (#2) #3.}
\def\rmp#1#2#3  {{\em Rev. Mod. Phys.} {\bf#1} (#2) #3.}
\def\rpp#1#2#3  {{\em Rep. Prog. Phys.} {\bf#1} (#2) #3.}
\def\sjnp#1#2#3 {{\em Sov. J. Nucl. Phys.} {\bf#1} (#2) #3.}
\def\spj#1#2#3  {{\em Sov. Phys. JEPT} {\bf#1} (#2) #3.}
\def\spu#1#2#3  {{\em Sov. Phys.-Usp.} {\bf#1} (#2) #3.}
\def\zp#1#2#3   {{\em Zeit. Phys.} {\bf#1} (#2) #3.}

\title{The Future of Top Physics at the Tevatron and LHC}

\author{Tony M. Liss\\
Department of Physics, University of Illinois\\
Urbana, IL 61801 }
\date{}

\begin{document}
\maketitle

\section{Introduction}
With the recent discovery of the top quark at Fermilab,~\cite{CDF,D0}
top  physics has moved from the search phase into the study phase.
The very large mass of the top quark separates it from the other fermions
and presents the possibility that new
physics may be discovered in either its production or its decays.

In this paper I review the prospects for measurements of the top quark
production and  decay parameters over the course of the next ten years or so.

\section{Top Yields}
\subsection{Tevatron Accelerator Upgrades}
With the turn on of the Main Injector at the Tevatron
it is predicted that
the average  instantaneous luminosity will reach $2\times10^{32}~cm^{-2}s^{-1}$
with a peak  luminosity of $5\times10^{32}~cm^{-2}s^{-1}$.  Integrated
luminosity delivered during Run II, beginning in 1999, is expected to be 2
fb$^{-1}$.  In addition to these luminosity  upgrades, the Tevatron will also
be
undergoing an energy upgrade from $\sqrt{s}=$1.8  TeV, to $\sqrt{s}=2.0$ TeV,
which gives an approximate 40\% increase in the $\ttbar$  production cross
section.

\subsection{Tevatron Detector Upgrades}
CDF and D0 both have significant detector upgrades planned prior to Run II.  I
here use  CDF as an example to calculate $\ttbar$ yields, but the numbers for
D0
should be  similar.

Significant tracking upgrades are planned at CDF including a 3-D silicon
tracker
and a  fiber tracker which will allow stand-alone tracking out to\\
$\mid\eta\mid$=2.0.  The  efficiency for tagging at least one b-jet in a
$\ttbar$ event is expected to be 80\% and for  double tagging close to 40\%.

Improvements in lepton identification will come from an upgrade to the end plug
calorimeter and through the completion of the muon coverage.  The increase in
the  acceptance for electrons from $\ttbar$ decays is expected to be 36\% and
that for  muons 25\%.

Including the 40\% increase in $\sigma_{\ttbar}$, an integrated luminosity of 2
fb$^{-1}$  will yield approximately 1400 {\em tagged} W plus $\geq$3 jet events
and about 140  dilepton events from $\ttbar$ decays, per experiment.

\subsection{Yields at the LHC}
Top physics at the LHC is expected to be done primarily during the early
running
at  relatively low luminosities of  $10^{32}-10^{33}~cm^{-2}s^{-1}$.  At
$\sqrt{s}$=14 TeV and $10^{32}~cm^{-2}s^{-1}$, this  corresponds to about 6000
$\ttbar$ pairs produced per day.  Folding in typical detection  efficiencies,
one can expect of order 100 tagged lepton + jet events on tape per day   and
about 20 dilepton events on tape  per day.  Only searches for the rarest
phenomena in top production or decays will be  limited by statistics at the
LHC.

\section{Mass Measurement}
\subsection{Current Status and Prospects at the Tevatron}
The current CDF top mass measurement, from an integrated luminosity of 67
pb$^{-1}$  is $M_{top}=176\pm 8\pm10$ GeV/c$^2$, where the first uncertainty is
statistical and  the second is systematic.  One observes that already the
statistical and systematic  uncertainties are comparable and therefore the
future precision will be determined by systematic effects (for Run II the
statistical uncertainty will be $\sim$1 GeV/c$^2$).

Both CDF and D0 use constrained fitting of b-tagged W+$\geq$ 4 jet events,
using
the  four highest $\Et$ jets, to measure the mass.  The dominant systematic
effects are due  to the understanding of the jet energy scale in the detector,
the effects of gluon  radiation, biases due to b-tagging, and the understanding
of the shape of the underlying  background.  In evaluating how the systematic
uncertainties are likely to scale with  increased integrated luminosity, the
key
question is whether a control dataset exists with  which to study the effect in
question.  If so, then it is reasonable to assume that the  uncertainty will
scale as $1/\sqrt{N}$.

Uncertainties due to jet energy scale and gluon radiation effects are studied
using  photon-jet balancing and Z + jet events, thus energy scale uncertainties
can be  expected to scale down as $1/\sqrt{N}$.
However, in
addition to energy scale  uncertainties, the effects of initial and final state
radiation create combinatoric confusion.
Combinatoric effects can be significantly reduced with increased statistics by
such  things as requiring both b-jets to be tagged, which reduces the number of
possible  particle assignments to four,
and  through the requirement of four and only 4 high $\Et$ jets
in the event.  It is unclear how  uncertainties due to combinatoric effects
will
scale, but it is clear that they will be  significantly reduced with a large
increase in dataset size.

Uncertainties due to b-tagging bias are currently understood with control
samples of  inclusive lepton events and with the $\ttbar$ Monte Carlo samples.
This uncertainty is  not large to begin with and should scale statistically.

The uncertainty due to the background shape is currently studied using the
VECBOS  Monte Carlo program, which is in turn validated using
W+1,2 jet data.   With increased dataset size it should be possible to study
the
background shape using top depleted datasets, selected for instance by {\em
anti}-b tagging, and with Z+$\geq$ 3 jet events.
It is conceivable that  background shape uncertainties will
scale statistically, but uncertain.

If we assume statistical scaling of the systematic uncertainties from the
current 67  pb$^{-1}$ values, the uncertainty projected at the completion of
Run
II will be 2-3  GeV/c$^2$.  Conservatively we can expect $< 4$ GeV/c$^2$.

\subsection{Prospects at LHC}
The subject of top mass fitting at LHC has been studied by several
authors.~\cite{Unal,Froidevaux,Mekki}  In the lepton plus jets channel, the
technique will likely be quite similar to that now employed at the Tevatron.
In the LHC event samples, the statistical uncertainty
in the mass fitting will be negligible, and the systematic effects
will be similar to those now under study at the Tevatron.  With the enormous
datasets, however, several new handles for control of the systematics are
available.  We list below the major systematic effects and the available
datasets used to control the uncertainties:
\begin{itemize}
\item{Jet Energy Scale:}  With large enough statistics the relevant jet energy
scale can be measured directly in $\ttbar$ events by renormalization of the
reconstructed mass of the hadronically decaying W boson.  Indeed, at CDF
one is already able to reconstruct a W mass peak on the hadronic side in
$\ttbar$ events when the W mass constraint is {\em removed} from the fitting
procedure.~\cite{lep-pho}  The difference between the energy scale for jets
from the W decay and the energy scale for b-jets will become more important
as the uncertainty in the mass measurement decreases.  With sufficiently
large datasets, the b-jet energy scale can be understood using
${\rm Z}\rightarrow\bbbar$ and ${\rm WZ}\rightarrow\ell\nu\bbbar$ events.
\item{Gluon Radiation:}  The LHC will provide copious samples of Z+jet
events which provide an excellent sample for measuring the effect of gluon
radiation on jet reconstruction.  Furthermore, it may be possible to study
hard gluon radiation, which can produce additional high $\Pt$ jets, in $\ttbar$
events themselves by measuring the population of additional jets.
\item{b-Tagging Bias:} This systematic effect is already small at the Tevatron
and is expected to be negligible at the LHC.
\item{Background Shape:} There is little work on the effect of the uncertainty
in the shape of the underlying background on the top mass fitting at the LHC,
but it should be possible to study this directly using carefully constructed
`top-free' datasets.
\end{itemize}
The LHC literature quotes an overall uncertainty on the top mass measurement
from lepton plus jets events of $\pm$3 GeV/c$^2$.  This work was done prior
to the discovery of top at the Tevatron, and given the current experience seems
extremely conservative.  More likely the final uncertainty at LHC will be in
the 1-2 GeV/c$^2$ range.

\section{Measurement of $\sigma_{\ttbar}$}
The measurement of the production cross section for $\ttbar$ pairs is a test of
QCD.  A significant deviation of the measured cross section from the predicted
value can signal non-Standard Model production mechanisms such as the decay of
a
heavy object into $\ttbar$ pairs.  As there is relatively little uncertainty in
the theoretical prediction for the cross section,~\cite{Berger} this can be a
rather sensitive testing ground.  The Tevatron and LHC measurements of
$\sigma_{\ttbar}$ are complementary. Although the LHC at higher $\sqrt{s}$
has greater reach, the dominant glue-glue nature of its
collisions makes it insensitive to a spin one color singlet resonance, while
there is no such restriction at the Tevatron where high $\sqrt{\hat{s}}$
collisions are dominantly $q\bar{q}$.

The uncertainty in the production cross section at the Tevatron is currently
of order 30\% and is dominated by statistics.  With the statistics of Run II,
one can expect a 10\% measurement.  At this point uncertainties due to
acceptance and integrated luminosity become comparable to the statistical
uncertainty and it is uncertain how much more precise the measurement can
become.  In any case, the ultimate precision in the luminosity is 3.5\%, which
is the accuracy to which the effective cross section of the luminosity monitors
is known, so one can expect that the uncertainty in the cross section
measurement will plateau in the range 5-10\% .  The final uncertainty in the
LHC measurement is likely to be in the same range.

\section{Single Top Production}

Single top production can occur through both the W-gluon fusion process, with a
t-channel W \cite{Yuan} or through an s-channel W$^*$
decay.~\cite{Willenbrock}
In either case the final state contains one top and one bottom quark and, in
lowest order, nothing else in the case of s-channel W$^*$ decay, while the
W-gluon fusion process contains an additional light quark jet in the final
state.  The cross section for single top production is proportional to the
square of the CKM element V$_{tb}$ and is  therefore of great interest.

\subsection{W-gluon Fusion}
The signal for single top production is extracted via the Wb invariant mass
distribution in b-tagged events.  There is a significant
background from $\qqbar\rightarrow\ W\bbbar$ and the signal to background at
the Tevatron is expected to be only 1:2.  With 2 fb$^{-1}$ of Run II data
however, the Wb mass peak can be extracted above background and a cross section
measurement with a statistical uncertainty of better than 20\% is
expected.~\cite{tev2000}  Extraction of V$_{tb}$ from this measured cross
section depends on the knowledge of the gluon distribution function and
therefore an accuracy of no better than 30\% is expected.  Comparable S/B and
final uncertainty on V$_{tb}$ can be expected at LHC.

\subsection{W$^*\rightarrow$tb}
The single top signal in this decay mode is also extracted via the Wb invariant
mass distribution, as above.  However, the backgrounds from the W-gluon fusion
process can be reduced by vetoing events with
additional jets and from W$\bbbar$ by making an invariant mass cut on
the $\bbbar$ pair of $>$110
GeV/c$^2$ (this cut is more efficient for the W$^*$ process
than for W-gluon fusion).~\cite{Willenbrock}

This process has a significant advantage over the W-gluon fusion process
for measuring V$_{tb}$ because it is not sensitive to the gluon distribution
function, and the uncertainties in the quark distributions can be controlled
by normalizing to $\qqbar\rightarrow\ell\nu$ at the same $\sqrt{\hat{s}}$.
With
2 fb$^{-1}$ at the Tevatron, a 12\% measurement of V$_{tb}$ is expected.

At LHC the W$^*$ signal is swamped by W-gluon fusion and $\ttbar$ production,
both of which grow faster with $\sqrt{s}$ than the $\qqbar$ initiated W$^*$
process. It
is unlikely therefore, that the measurement of V$_{tb}$ with this technique
at LHC will compete with the measurement at the Tevatron.

\section{Measurement of V$_{tb}$ from Top Decays}
In addition to the single top measurements discussed above, there is also
sensitivity to V$_{tb}$ by measuring the branching fraction $t\rightarrow
Wb$/$t\rightarrow Wq$.  This branching fraction is currently measured at CDF,
via the ratio of $\ttbar$ events with 0,1 or 2 b-tagged jets, to about 30\%.
With 2 fb$^{-1}$ a branching fraction uncertainty of 10\% is expected.

Converting the branching fraction measurement to a measurement of V$_{tb}$
requires assumptions about the magnitudes of V$_{td}$ and V$_{ts}$.  Since
these latter two CKM elements are quite small in the Standard Model, the
branching fraction measurement is not a terribly sensitive way to measure
V$_{tb}$, assuming V$_{tb}$ is close to 1 .  A 10\% measurement of the
branching fraction corresponds to a 1$\sigma$ lower limit on V$_{tb}$ of
0.26.  At LHC, assuming a branching fraction uncertainty of 1\%, the lower
limit on V$_{tb}$ is only 0.4 .

\section{$t\rightarrow H^+b$}
Supersymmetric models include charged Higgs bosons which couple to top quarks
with a strength which depends on the Higgs mass and the ratio of vacuum
expectation values, $\tan\beta$.  The Higgs subsequently decays to $\tau\nu$ or
$cs$. The branching fraction dependence of the top and Higgs decays on
$\tan\beta$ is shown in Figure 1.

The ratio, R$_{\ell\ell}$,  of the rate of $\ttbar$ pairs decaying into
dilepton
final states to those decaying into single lepton final states is, in
principle,
sensitive to the presence of a charged Higgs component of the top decays.
However, the contributions to both the dilepton and single lepton final states
from $H\rightarrow\tau\nu,~cs$ must first be understood.

Extrapolating from the current Tevatron experience, R$_{\ell\ell}$ will be
measured to $\sim$10\% with 2 fb$^{-1}$, where the uncertainty is dominated by
the statistics of the dilepton sample.  At LHC, the measurement is dominated
by uncertainties in the backgrounds, which can be controlled somewhat via
b-tagging in both the single lepton and dilepton channels.~\cite{Unal}  A
measurement of the ratio to less than 5\% is a reasonable expectation.
In either case, Tevatron
or LHC, there is sensitivity to only a limited $\tan\beta$ range for a
given Higgs mass (see Fig.1).

A more direct method for searching for charged Higgs decays is to look for an
excess of taus in top decays.  An LHC study~\cite{Felcini} has shown that this
technique can be sensitive to a charged Higgs over most of the $\tan\beta$
range
if M$_{Higgs}$ is near 150 GeV/c$^2$.

\section{W-t-b Vetex}
The most general form of the W-t-b vertex has 4 form factors: $F_L^1,F_R^1,
F_L^2,~{\rm and}~F_R^2$.~\cite{Kane}  In the Standard Model, only $F_L^1$, the
V-A form factor, is
non-zero at lowest order.  Experimental sensitivity to the values of these form
factors can be achieved by measuring the polarization of the W bosons in top
decays.  In the Standard Model, the ratio of the number of longitudinally
polarized to transversely polarized W bosons depends on M$_{top}^2$ and gives
about 70\% longitudinally polarized
Ws for M$_{top}$=175 GeV/c$^2$.  The fraction of longitudinal W bosons can
be measured using the angular distribution of leptons from W decays in top
events.  Non-zero values of $F^2_{L,R}$ would produce
a departure from the predicted value.  A non-zero $F^1_R$ gives a
V+A component of the coupling but does not affect the fraction of longitudinal
W bosons.  However, it would produce a right-handed component in the transverse
Ws and therefore there is also sensitivity to $F^1_R$ in the lepton angular
distributions.

Studies at the Tevatron have shown that a 2 fb$^{-1}$ sample would give a
statistical uncertainty of
3\% on the fraction of longitudinal, and a
1\% statistical uncertainty on the fraction of right handed,
W bosons in top decays.
At LHC the statistical uncertainties will be a factor of 3-10 better.
In both cases systematic effects, which remain to be studied, are likely
to dominate the precision of the measurement.

\section{Rare Decays}
Flavor changing neutral current decays such as $t\rightarrow\gamma c$ and
$t\rightarrow Zc$ are unobservably small in the Standard Model at either the
Tevatron or LHC.  Any observation of such decays at either machine would be a
breakthrough.  With 2 fb$^{-1}$ at the Tevatron, branching fraction limits for
either of these decays will be at the per cent level, whereas the Standard
Model
predictions are eight orders of magnitude smaller. The Standard Model
prediction
for the decay $t\rightarrow Ws$ is of order $10^{-3}$, which is about an order
of magnitude smaller than the sensitivity at the Tevatron with 2 fb$^{-1}$.  It
is possible that LHC will make the first observation of this rare decay.

\section{Conclusions}
By the turn of the century, Run II at the Tevatron will have produced 2
fb$^{-1}$ of data for both CDF and D0.  Each experiment will have measurements
of the top mass to better than 4 GeV/c$^2$, the production cross section to
10\%, V$_{tb}$ to 12\% as well as searches for rare decay modes of the top. The
charged current couplings of the top will have been probed to a few per cent
via
W polarization measurements.  If, as many hope and some expect, the top quark
turns out to be a window onto physics beyond the Standard Model, these
measurements at the Tevatron may very well yield the first glimpse of that new
physics.

R\&D efforts at Fermilab are now under way to evaluate the
possibility of even higher luminosities. Running at 10$^{33}$ cm$^{-2}$s$^{-1}$
or a constant $5\times 10^{32}$ cm$^{-2}$s$^{-1}$ is considered possible,
and might yield as much as 10 fb$^{-1}$ per year.

In a year of LHC running at $10^{32}-10^{33}$cm$^{-2}$s$^{-1}$, $\ttbar$
samples
will be 1-2 orders of magnitude larger than those with 2 fb$^{-1}$ at the
Tevatron.  With such samples, many LHC measurements will be limited by
systematic effects which are difficult to quantify at this point.
Nevertheless,
improvements by a factor of 2-3 in the top mass uncertainty, and at least that
much in sensitivity to rare decays and non-standard charged current couplings
seem reasonable expectations.  While Tevatron running should produce the best
measure of V$_{tb}$, the first observation of $t\rightarrow Ws$ may come from
LHC.  If new physics does show up in top production or decays, then LHC is in
an
excellent position to either discover it, or study it if it should be found
first at Fermilab.

\section{Acknowledgements}
This paper draws on the work of many other people, hopefully covered adequately
in the references.  In particular, I would like to thank D. Amidei,
D. Froidevaux, and S. Willenbrock for helpful discussions during
its preparation.

\newpage

\end{document}